\documentclass[epj]{webofc}
\usepackage[varg]{txfonts}   
\usepackage{hyperref}

\renewcommand*{\eqref}[1]{Eq.~(\ref{eq:#1})}

\newcommand*{\figref}[1]{Fig.~(\ref{fig:#1})}
\newcommand*{\figlab}[1]{\label{fig:#1}}

\newcommand*{\seclab}[1]{\label{sec:#1}}
\wocname{\includegraphics[width=0.25cm,clip]{logoARENA} ARENA2018}
\wocname{ARENA2018}
\woctitle{ARENA2018}
\woctitle{\includegraphics[width=0.25cm,clip]{logoARENA} ARENA2018}
\begin{document}
\title{Signal recognition and background suppression by matched filters and neural networks for Tunka-Rex
}
%
%

\author{
\firstname{D.} \lastname{Shipilov}\inst{1},
\firstname{P.A.} \lastname{Bezyazeekov}\inst{1},
\firstname{N.M.} \lastname{Budnev}\inst{1},
\firstname{D.} \lastname{Chernykh}\inst{1},
\firstname{O.} \lastname{Fedorov}\inst{1},
\firstname{O.A.} \lastname{Gress}\inst{1},
\firstname{A.}~\lastname{Haungs}\inst{2},
\firstname{R.} \lastname{Hiller}\inst{2}\fnsep\thanks{now at the University of Zürich},
\firstname{T.} \lastname{Huege}\inst{2}\fnsep\thanks{also at Vrije Universiteit Brussel, Brussels, Belgium},
\firstname{Y.} \lastname{Kazarina}\inst{1},
\firstname{M.} \lastname{Kleifges}\inst{3},
\firstname{E.E.}~\lastname{Korosteleva}\inst{4},
\firstname{D.}~\lastname{Kostunin}\inst{2},
\firstname{L.A.} \lastname{Kuzmichev}\inst{4},
\firstname{V.} \lastname{Lenok}\inst{2},
\firstname{N.} \lastname{Lubsandorzhiev}\inst{4},
\firstname{T.} \lastname{Marshalkina}\inst{1},
\firstname{R.}~\lastname{Monkhoev}\inst{1},
\firstname{E.} \lastname{Osipova}\inst{4},
\firstname{A.} \lastname{Pakhorukov}\inst{1},
\firstname{L.} \lastname{Pankov}\inst{1},
\firstname{V.V.} \lastname{Prosin}\inst{4},
\firstname{F.G.} \lastname{Schröder}\inst{2,}\inst{5}\and
\firstname{A.}~\lastname{Zagorodnikov}\inst{1}
(Tunka-Rex Collaboration)
}


\institute{Institute of Applied Physics ISU, Irkutsk, Russia \and
Institut für Kernphysik, Karlsruhe Institute of Technology (KIT), Karlsruhe, Germany \and
Institut für Prozessdatenverarbeitung und Elektronik, Karlsruhe Institute of Technology (KIT), Karlsruhe, Germany \and
Skobeltsyn Institute of Nuclear Physics MSU, Moscow, Russia \and
Department of Physics and Astronomy, University of Delaware, Newark, DE, USA
          }

\abstract{%
The Tunka Radio Extension (Tunka-Rex) is a digital antenna array, which measures radio emission
of the cosmic-ray air-showers in the frequency band of 30-80~MHz. Tunka-Rex is co-located with the TAIGA experiment in Siberia and consists of 63 antennas, 
57 of them are in a densely instrumented area of about 1 km\textsuperscript{2}. 
In the present work we discuss the improvements of the signal reconstruction applied for Tunka-Rex. 
At the first stage we implemented matched filtering using averaged signals as template. 
The simulation study has shown that matched filtering allows one to decrease the threshold
of signal detection and increase its purity. However, the maximum performance of matched
filtering is achievable only in case of white noise, while in reality the noise is not fully random due
to different reasons. To recognize hidden features of the noise and treat them, we decided to use
convolutional neural network with autoencoder architecture. Taking the recorded trace as an
input, the autoencoder returns denoised traces, i.e. removes all signal-unrelated amplitudes. 
We present the comparison between the standard method of signal reconstruction, matched filtering and the autoencoder, and discuss
the prospects of application of neural networks for lowering the threshold of digital antenna arrays
for cosmic-ray detection.}
\maketitle
\section{Introduction}
\label{intro}
Tunka-Rex is an antenna array, which measures the radio emission of air showers produced by cosmic rays with energies above 100 PeV in the frequency band of 30-80 MHz \cite{Bezyazeekov:2015rpa}. Tunka-Rex requires an external trigger and operates jointly with the non-imaging air-Cherenkov light detector Tunka-133 \cite{Prosin:2014dxa} and the  scintillators  of  Tunka-Grande \cite{Budnev:2015cha}.

The main background at the Tunka-Rex location is the Galaxy. 
However there are many sources of non-white and non-stationary background in the Tunka Valley. 
Due to this we use two different approaches: matched filter with predefined signal template and neural network with optimized convolutional filters.
\section{Matched filter and autoencoder for signal reconstruction}
\seclab{sec-1}
In the present work we use 650\,000 samples of measured Tunka background and 25\,000 CoREAS simulations folded with Tunka-Rex hardware response. We use single polarization ($v\times B$) and the following upsampling rates: 64 for matched filter and 16 for neural network. 
The true peak position in the simulated trace is defined as a position of the maximum of the Hilbert envelope of the trace before adding the background~\cite{Bezyazeekov:2015rpa}.

\begin{figure}[t!] 
  \label{fig-1} 
  \begin{minipage}[b]{0.40\linewidth}
    \centering
    \includegraphics[width=1.0\linewidth]{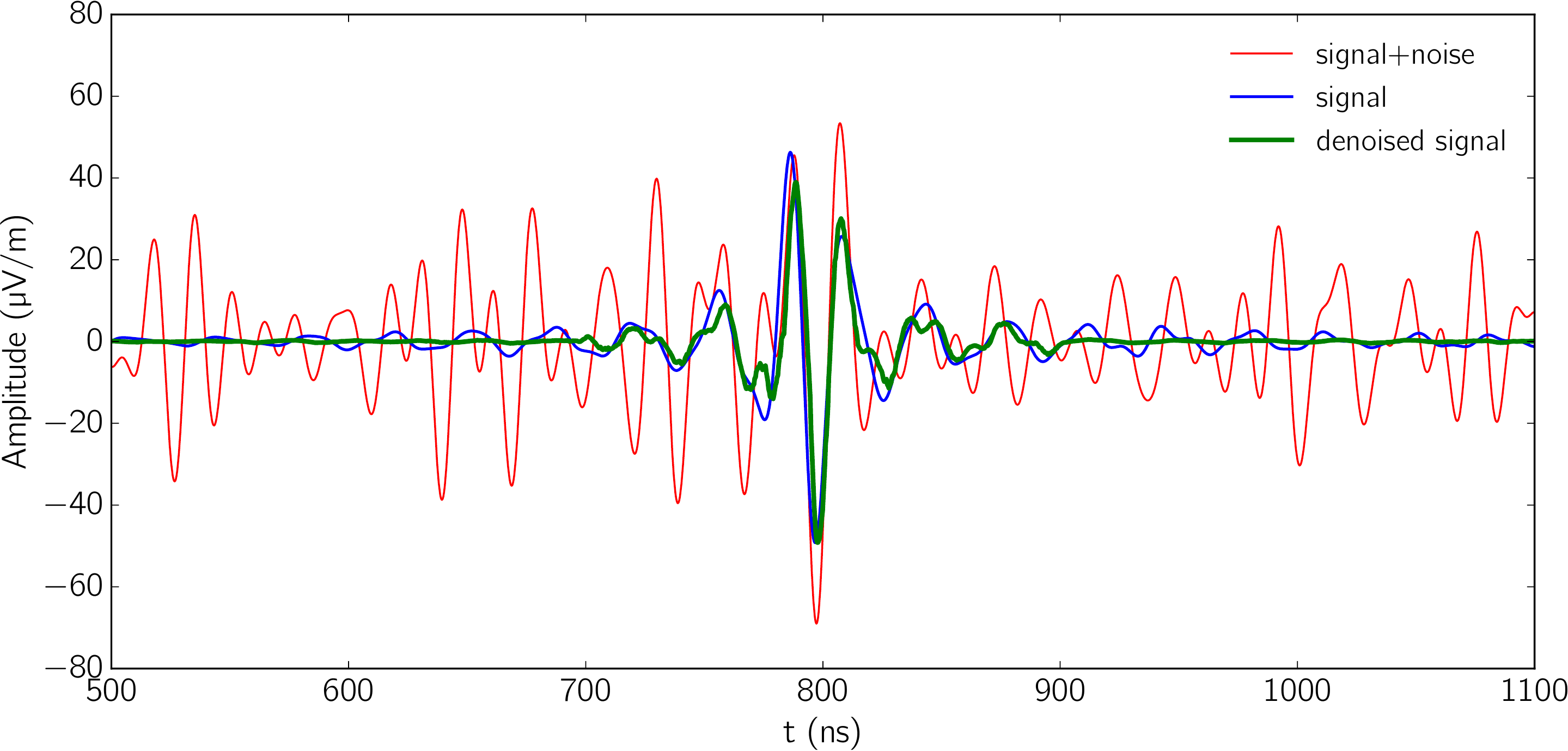} 
     \includegraphics[width=1.0\linewidth]{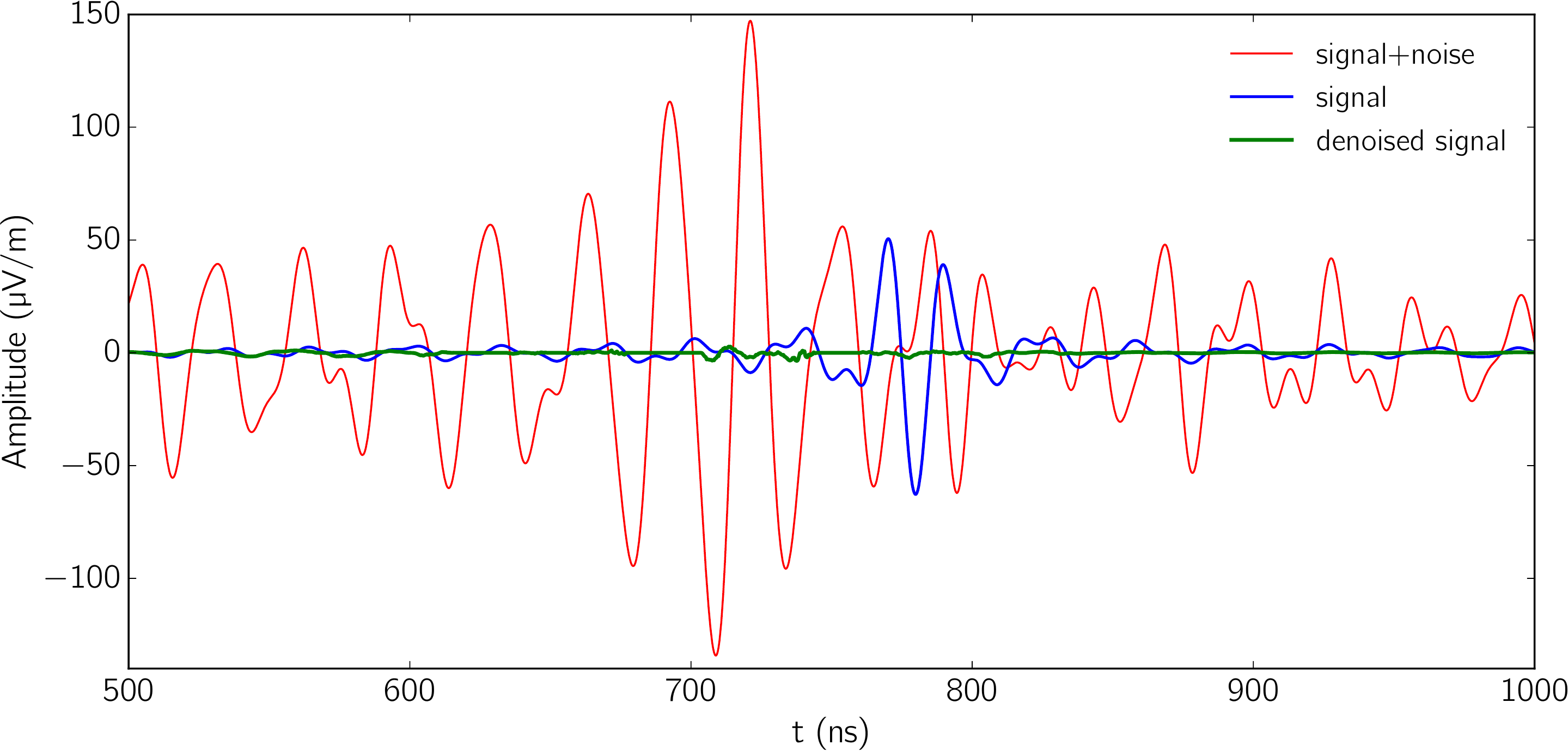} 
     \includegraphics[width=1.0\linewidth]{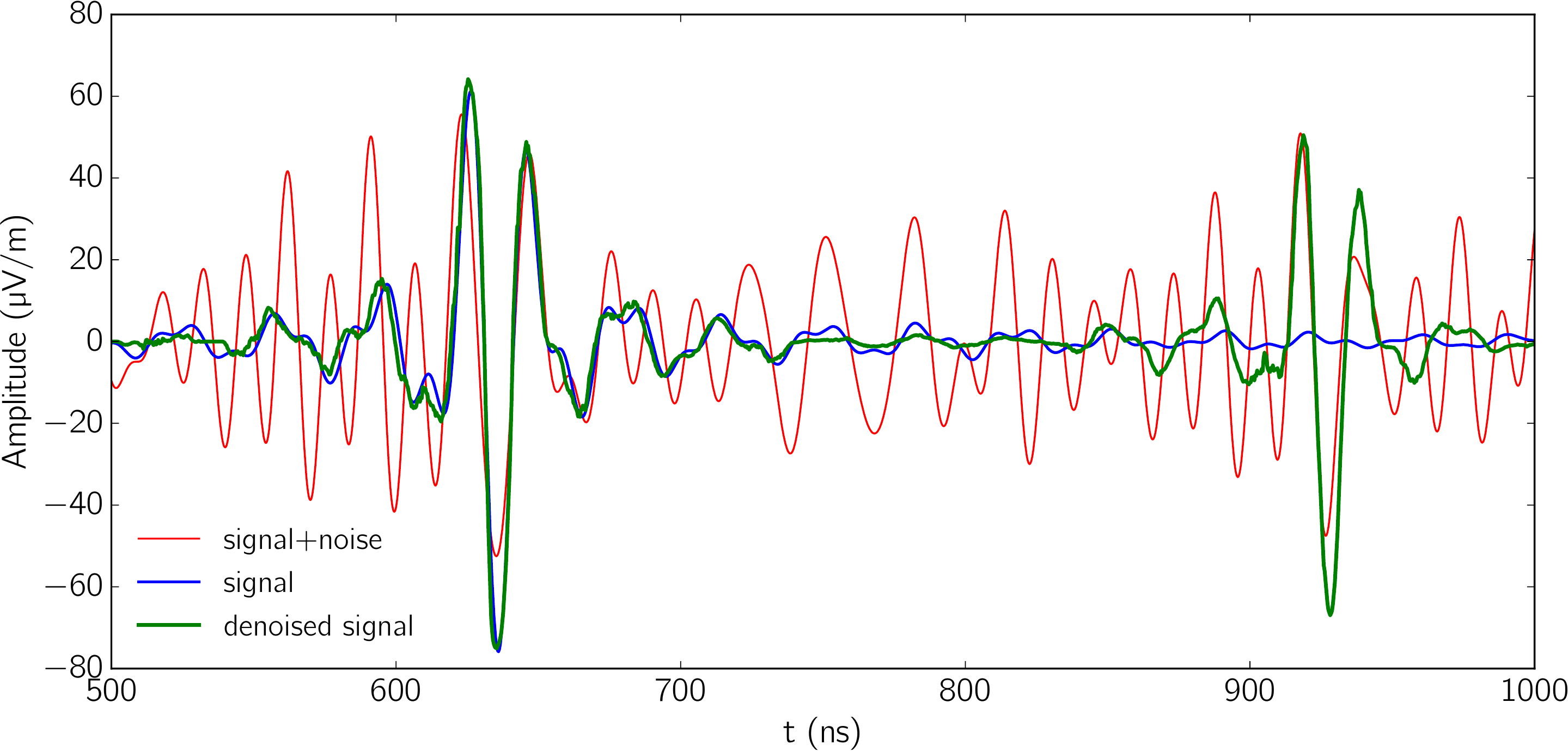} 
  \end{minipage} 
   \begin{minipage}[b]{0.59\linewidth}
   \centering
    \includegraphics[width=0.45\linewidth]{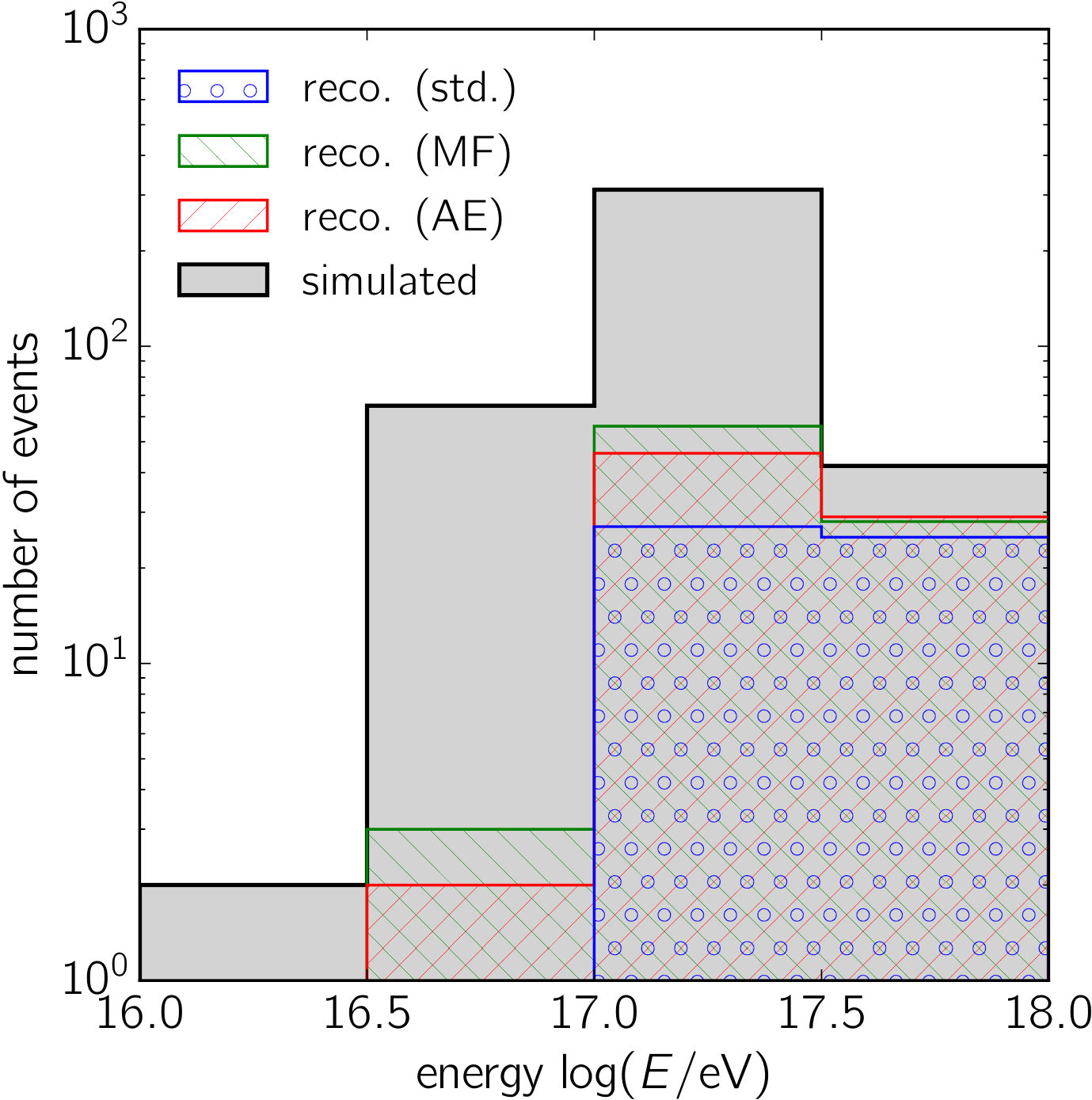} ~
    \includegraphics[width=0.45\linewidth]{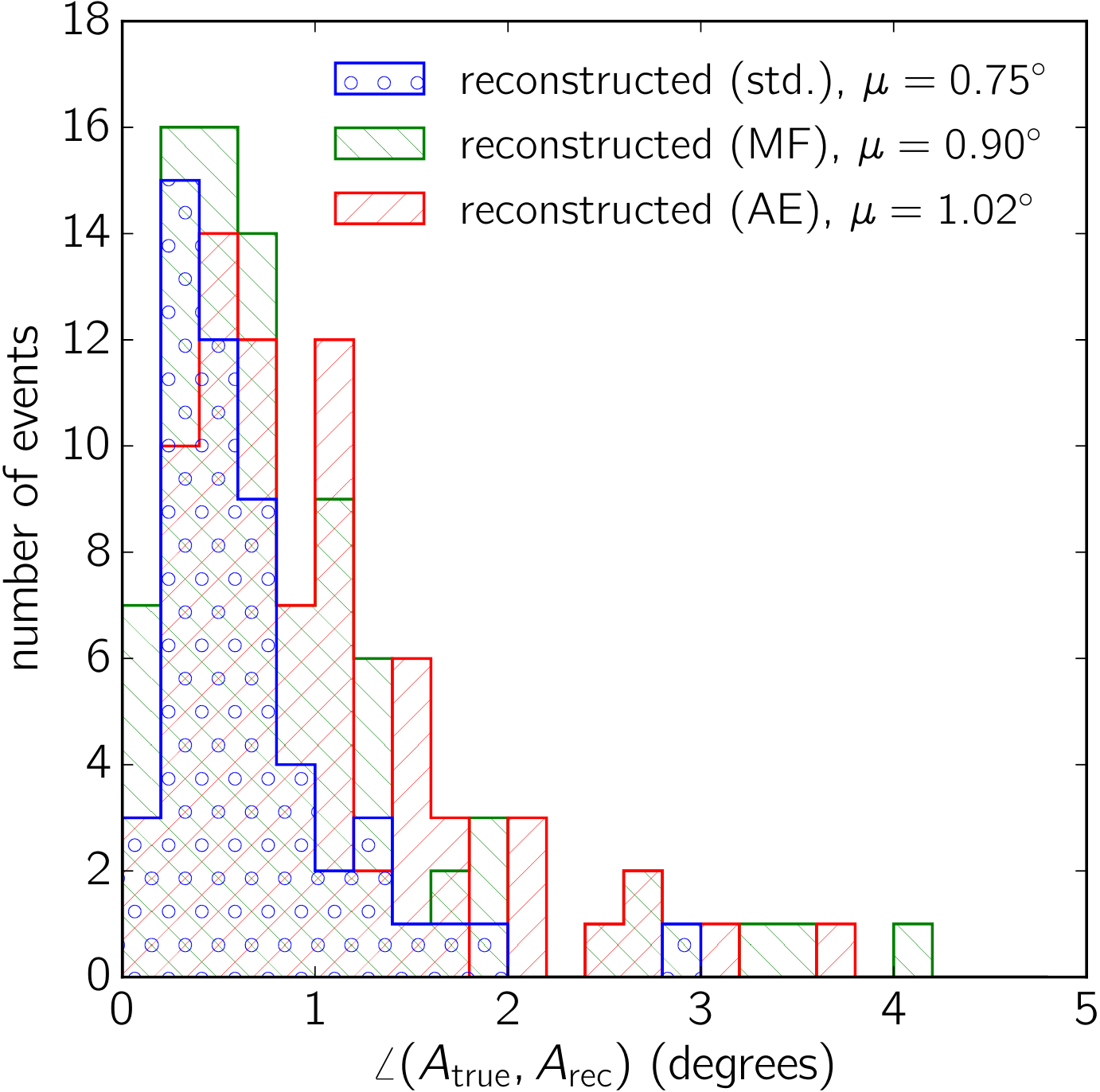} 
   ~\\
      ~\\
    \includegraphics[width=0.45\linewidth]{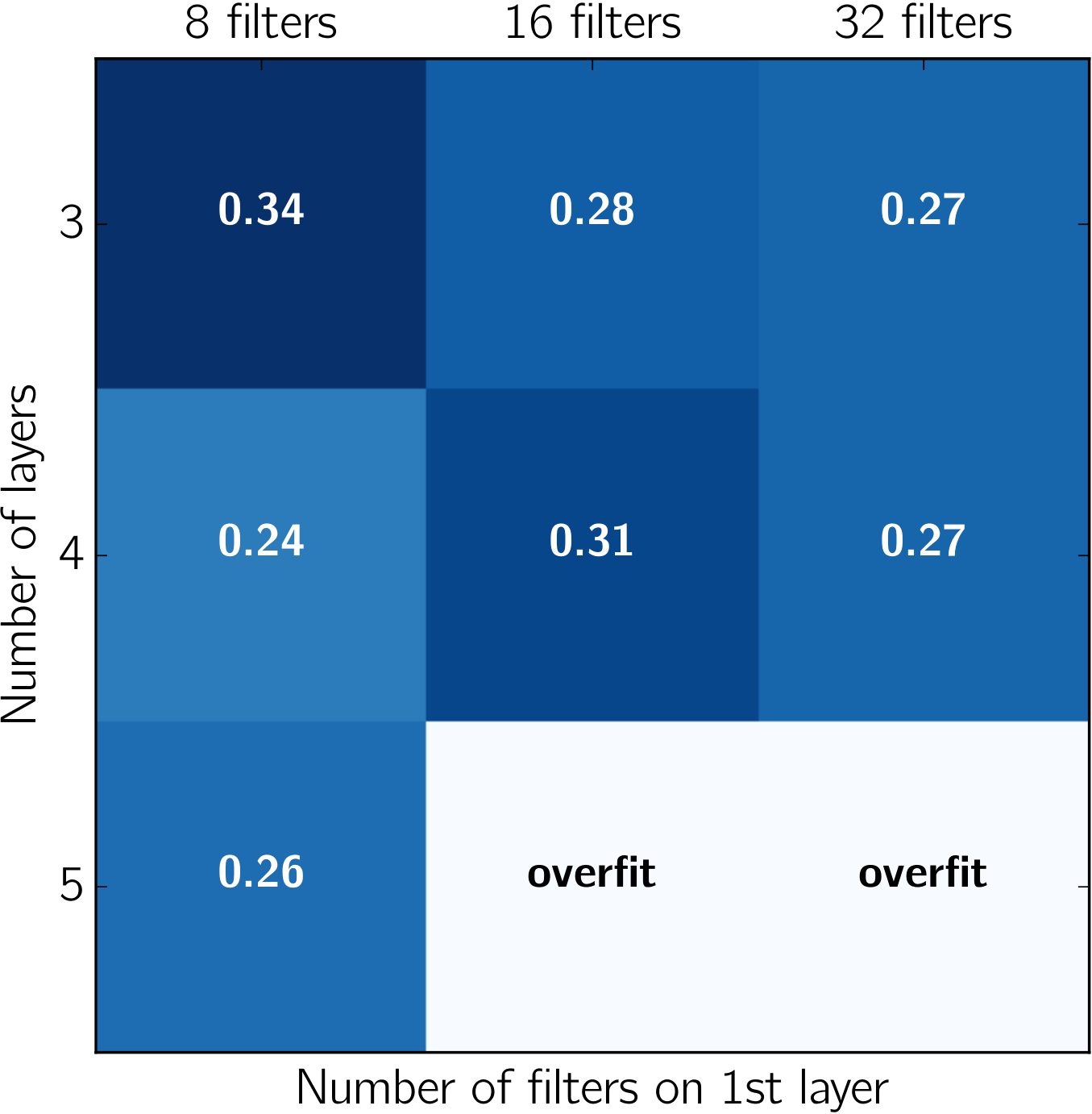} ~
    \includegraphics[width=0.45\linewidth]{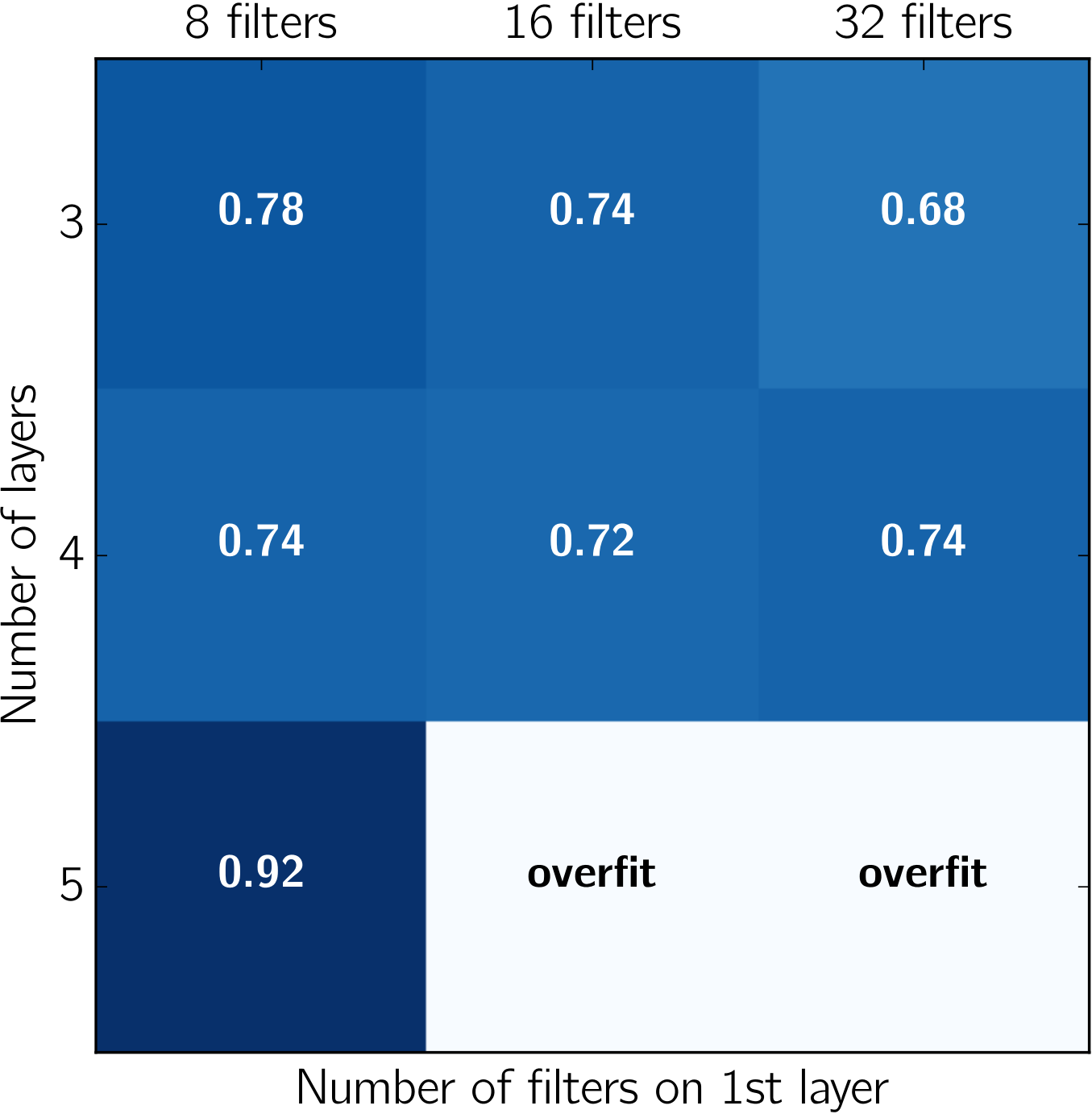} 
  \end{minipage} 
\caption{Left panel, examples of AE performance: correct identification (true positive), no identification (true negative), double identification (true plus false positive). Right panel, from the left upper conner to the bottom right:  1) distribution of reconstructed events as function of energy; 2) angular resolution of
different methods; 3) efficiency of AE; 4) purity of AE. }
\figlab{fig-1}      
\end{figure}

A matched filter (MF) convolutes a template with the input trace and the maximum of the convolution defines the position of the peak. Templates are obtained from averaging of many CoREAS\cite{Huege:2013pro} simulations. In the present work we use templates with lengths of 60 ns, upsampled to 800 MS/s rate (see \figref{fig-1}).
The threshold is defined as 5\% probability of false positive. 
Amplitude is estimated as the function of the square root of cross-correlation. 
We have implemented the MF in the Auger Offline \cite{Abreu:2011pro}\ and tested it on the set of simulated events. 
The MF is able to reconstruct pulses with lower amplitudes and features resolution of arrival direction similar to standard method. The distribution of  reconstructed events and arrival directions can be seen in \figref{fig-1}.
Here it is worth noticing, that we did not perform any corrections and whitening of our noise, by this the performance of the MF does not reach it full efficiency.

At the next step we use neural network which called autoencoder (AE). 
AE is based on 1D convolutional layers with rectified linear units and max pooling after convolution layers. Binary cross-entropy is used as a loss function. 
For minimization of the loss, all data should be normalized in [0;1] range and baseline always should be put on a 0.5 level, what helps the AE to extract features from noise.

The structure of the AE is defined by the following: depth ($D$) and number of filters per layer ($N$) are free parameters. The $i$-th encoding layer ($i=1,...,D$) is described by the following:
\begin{equation}
S_i = S_{\mathrm{min}} \times 2^{D-i}\,,\,\,\, n_i = 2^{i + N-1}\,,
\end{equation}
where $S_i$ is size of the $i$-th filter, $n_i$ is a number of filters per layer. $S_\mathrm{min} = 16$ is minimal size of layer (corresponding to few ns).

To assess the quality of the networks we have introduced 2 metrics: efficiency: $N_\mathrm{rec.}/N_\mathrm{tot.}$, namely fraction of events that passed the threshold; and the purity: $N_\mathrm{hit}/N_\mathrm{rec.}$, namely the fraction of events with reconstructed position of the peak $|t_\mathrm{rec.} - t_\mathrm{true}|<5$~ns.
Networks with 3-4 layers show similar result, only addition of the 5th layer increases of the purity. 
Networks with 5 layers have a large number of degrees of freedom which leads to overfitting with our limited size of the training dataset. \\
\section{Summary}
\seclab{sec-4}
We have improved Tunka-Rex signal reconstruction by implementing matched filters and autoencoders. New methods show promising results in the lowering of the threshold. We will improve current results by extending the library of templates for matched filter and optimization of autoencoder architectures.
\section*{Acknowledgements} 
This work is supported by the Helmholtz grant HRSF-0027, the Russian Federation Ministry of Education and Science (Tunka shared core facilities, unique identifier RFMEFI59317X0005, 3.9678.2017/8.9, 3.904.2017/4.6, 3.6787.2017/7.8, 3.6790.2017/7.8), the Russian Foundation for Basic Research (Grants No. 16-02-00738, No. 17-02-00905, No. 18-32-00460). In preparation of this work we used calculations performed on the HPC-cluster “Academician V.M. Matrosov” and on the computational resource ForHLR II funded by the Ministry of Science, Research and the Arts Baden-W\"urttemberg and Deutsche Forschungsgemeinschaft. The work presented in Section 2 of this paper was funded by the Russian Science Foundation (the grant No. 18-41-06003).

\end{document}